 \documentclass[twocolumn]{aastex6}
\shorttitle{Excess Galactic molecular absorption toward 3C~111}
\shortauthors{Tombesi et al.}

\begin{document}



\title{Excess Galactic molecular absorption toward the radio galaxy 3C~111}


\author{F. Tombesi$^{1,2,3}$, C.~S. Reynolds$^2$, R.~F. Mushotzky$^2$
  and E. Behar$^4$}
\affil{$^{1}$X-ray Astrophysics Laboratory, NASA/Goddard Space Flight Center, Greenbelt, MD 20771, USA; francesco.tombesi@nasa.gov}
\affil{$^{2}$Department of Astronomy, University of Maryland, College Park, MD 20742, USA; ftombesi@astro.umd.edu}
\affil{$^3$Department of Physics, University of Rome ``Tor Vergata'',
  Via della Ricerca Scientifica 1, I-00133 Rome, Italy; francesco.tombesi@roma2.infn.it}
\affil{$^4$Department of Physics, Technion 32000, Haifa 32000, Israel}



\begin{abstract}



We show the combined spectral analysis of \emph{Chandra}  high
energy transmission grating (HETG) and \emph{XMM-Newton} reflection grating spectrometer (RGS) observations of the broad-line radio galaxy 3C~111.
The source is known to show
excess neutral absorption with respect to the one estimated from 21~cm radio surveys of atomic H~I in the
Galaxy. However, previous works were not able to constrain the origin of such
absorber as local to our Milky Way or intrinsic to the source ($z =
0.0485$). The high signal-to-noise grating spectra allow us to constrain the excess absorption as due to intervening gas
in the Milky Way, and we estimate a time averaged total column density of $N_H =
(7.4\pm0.1)\times 10^{21}$~cm$^{-2}$, a factor of two higher than
the tabulated H~I value. We recommend to use the
total average Galactic column density here estimated when studying 3C~111.
The origin of the extra Galactic absorption of $N_H = 4.4\times 10^{21}$~cm$^{-2}$ is likely
due to molecular gas associated with the Taurus molecular cloud complex toward 3C~111,
which is our nearest star-forming region.  We also detect a weak (EW$=$$16\pm10$~eV) and narrow
(FWMH$<$5,500 km~s$^{-1}$, consistent with optical H$\alpha$) Fe
K$\alpha$ emission line at E$=$6.4~keV likely from the torus in the
central regions of 3C~111, and we place an upper limit on the column
density of a possible intrinsic warm absorber
of $N_H$$<$$2.5\times10^{20}$~cm$^{-2}$. These complexities make
3C~111 a very promising object for studying both the
intrinsic properties of this active radio galaxy and the Galactic
interstellar medium if used as a background source.  



\end{abstract}


\section{Introduction} 

Active galactic nuclei (AGN) are classified in two main categories as
radio-loud and radio-quiet AGN, depending on their radio luminosity with
respect to optical (Kellerman et al.~1989). The origin of this
dichotomy is still not known, but it is manifested by the presence of
powerful, often relativistic, radio jets in radio-loud AGN (e.g., Urry
\& Padovani 1995). On the other hand, radio-quiet AGN, most notably Seyfert
galaxies and quasars, show clear features of massive,
sub-relativistic winds (e.g., Tombesi et al.~2010a; Gofford et
al.~2015). Although,
winds have been recently reported in radio-loud sources as well (e.g.,
Reeves et al.~2009; Torresi et al.~2010, 2012; Tombesi et al.~2010b,
2014; Gofford et al.~2015). 

The X-ray band is particularly promising for investigating the
possible origin of such dichotomy because the radiation is emitted
close to the central supermassive black hole (SMBH), and therefore it
can retain information from both the accretion disk and intervening
absorption and emission in the regions surrounding the AGN or the host galaxy
interstellar medium (e.g. Sambruna et al.~1999, 2011; Ogle et
al.~2005; Lewis et al.~2005; Ballo et al.~2011; Braito et al.~2011;
Tombesi et al.~2011, 2016; Lohfink et al.~2015). Coordinated
observations in the X-ray, optical, and radio allowed also to study the connection between the disk, jet and winds in
some sources (e.g., Chatterjee et al.~2011; Tombesi et al.~2012; Lohfink et al.~2013). 

Here, we report on the analysis of a long 150~ks \emph{Chandra} HETG
observation of the broad-line radio galaxy (BLRG) 3C~111 ($z =
0.0485$), in combination with archived \emph{XMM-Newton} RGS spectra. This is the
third paper of this series, the previous two focused on
3C~390.3 and 3C~120, respectively (Tombesi et al.~2016, 2017). 
The radio galaxy
3C~111 is an X-ray bright BLRG and it is classified as a
Fanaroff--Riley type II (FRII) source with a double-lobe/single-jet
morphology (Linfield \& Perley 1984). The inner jet has been seen to
display superluminal motion (Vermeulen \& Cohen 1994; Chatterjee et
al.~2011). This source was also detected in the $\gamma$-ray band with
Fermi (Kataoka et al.~2011; Grandi et al.~2012). 

The main objective of the \emph{Chandra} HETG campaing was to study
  the possible X-ray warm absorbers in the brightest radio
  galaxies. However, we do not find significant evidence for an intrinsic soft X-ray ionized absorber in this
  source. The apparent lack of a warm absorber is puzzling due to the
  fact that 3C~111 shows powerful jets and accretion disk winds driven
  by the central supermassive black hole (e.g., Tombesi et al.~2010b;
  Chatterjee et al.~2011). Moreover, warm absorbers have been detected
in most bright radio galaxies (e.g., Reeves et al.~2009; Torresi et
al.~2010, 2012; Braito et al.~2011). On the other hand, this source is known to show an excess of
cold/neutral absorption with respect to the value estimated from 21~cm
radio surveys of atomic H~I. Previous X-ray studies have not been able
to constrain the origin of such absorber as local to our Milky Way or intrinsic to 3C~111 because the
redshift could not be constrained (e.g., Reynolds
et al.~1998; Lewis et al.~2005; Ballo et al.~2011; Tombesi et
al.~2013). 

We note that 3C~111, along with other bright AGN, has been used as an extragalactic background
radio source to explore complex atomic and molecular gas regions in the
Milky Way (e.g., Marscher, Moore \& Bania 1993; Moore \& Marscher
1995). In fact, 3C~111 is located at a relatively low latitude with respect to the
Galactic plane and it lies behind the giant Taurus 
molecular cloud, which is the nearest large star-forming region in our Galaxy (Ungerer et al.~1985;
Ungerechts \& Thaddeus 1987).

Several detailed studies of the structure in the atomic
component of the interstellar medium (ISM) have suggested that
complexities exist on scales as small as few tens of AU. The first
indication that such small-scale H~I structures may exist was reported by
Dieter, Welch, \& Romney (1976). Using VLBI techniques to measure changes in the visibility
function in the Galactic H~I absorption line toward 3C~147, they
proposed that the line variations were due to a small cloud with
dimension of about 70 AU and a density of $\simeq 10^5$
atoms~cm$^{-3}$.

Moreover, Faison et al.~(1998) used VLBA and the VLA to image Galactic H~I in absorption in the direction
of three other bright extragalactic sources, namely 3C~138, 2255$+$416, and CJ1~0404$+$768.
They suggested that the small-scale opacity structures seen toward 3C~138 and 2255$+$416
may be due to density variations, spin temperature variations,
velocity turbulence in the atomic gas, or a combination of these
effects. If the opacity variations are due to fluctuations in density,
then they would suggest clouds with high densities of  $\sim 10^6$~cm$^{-3}$ in the cold
neutral medium on $\sim$10~AU scales.

Besides atomic H~I, there has been evidence for small-scale structures
also in the diffuse molecular gas in our Galaxy. For instance, Moore
\& Marscher (1995) reported changes in the column density of
formaldehyde (H$_2$CO) toward the compact sources NRAO~150, 3C~111,
and BL Lac. They observed the three sources for over 3~yrs with the
VLA in the 6~cm H$_2$CO line. The motion of the sources due to
parallax as well as the proper motions of the absorbing gas caused the
relative line of sight through the molecular gas to the extragalactic
sources to change with time. They observed signifcant variations in
the molecular column density toward NRAO~150 and 3C~111, which
indicate structures on the scales of $\sim$10~AU and densities of
$\sim 10^6$ cm~$^{-3}$. Several other studies have been performed in order to map the
molecular OH and CO distribution in the Galaxy using compact
background continuum sources in the radio and mm-waves (e.g., Liszt \&
Lucas 1996, 1998).

In the X-ray band, compact X-ray binaries have provided
important information regarding the composition and ionization of the
Galactic ISM through absorption, although such sources are mostly
distributed along the galactic plane (e.g., Schulz et al. 2002; Gatuzz
et al.~2015). Some attempts to use bright extragalactic continuum sources
to explore our Galaxy multi-phase and multi-scale ISM
along several lines of sight have also been reported in the UV and X-ray bands,
for instance recently in the context of the origin of the \emph{Fermi}
bubbles (e.g., Fox et al. 2015; Nicastro et al.~2016;Bordoloi et al.~2017). 

A complementary method to explore the atomic and molecular hydrogen
content from many different sightlines in the Milky Way is provided by gamma-ray bursts (GRBs) afterglows in the
X-ray band. We note that the value
reported for the cold absorption toward 3C~111 is about a factor
of two higher than the upper limit reported by Willingdale et
al.~(2013) for GRB sightlines.

\section{Data analysis and results} 

We describe the analysis of the \emph{Chandra} HETG spectrum of the
broad-line radio galaxy 3C~111. It was observed on November 4th 2014
for a single exposure of 143~ks (ID 16219). The spectrum was extracted using the \emph{CIAO}
package v4.7. Only the first order dispersed
spectra were considered for both the Medium Energy Grating (MEG) and
High Energy Grating (HEG), and the $\pm$1 orders for each grating were
subsequently combined for each sequence. The background coun trate is
found to be negligible. The resultant spectra were binned to the full
width half maximum (FWHM) of their spectral resolution, which 
corresponds to $\Delta\lambda = 0.023$ \AA\, and $\Delta\lambda =
0.012$ \AA\, bins for MEG and HEG, respectively. The MEG and HEG
spectra were analyzed in the energy bands E$=$0.5--7~keV and
E$=$1--7.5~keV, respectively. The MEG and HEG count rates are
0.47~cts~s$^{-1}$ and 0.28~cts~s$^{-1}$, respectively. The spectral
analysis was performed using the software \emph{XSPEC} v.12.8.2
and the C-statistic was applied. We performed simultaneous fits of the
MEG and HEG spectra considering a free cross-normalization constant,
which resulted being always very close to unity. All parameters are given in
the source rest frame and the errors and limits are at the 90\% level if not
otherwise stated. Standard Solar abundances are assumed (Asplund et al.~2009).

  \begin{figure}[t!]
  \centering
   \includegraphics[width=8.5cm,height=10cm,angle=0]{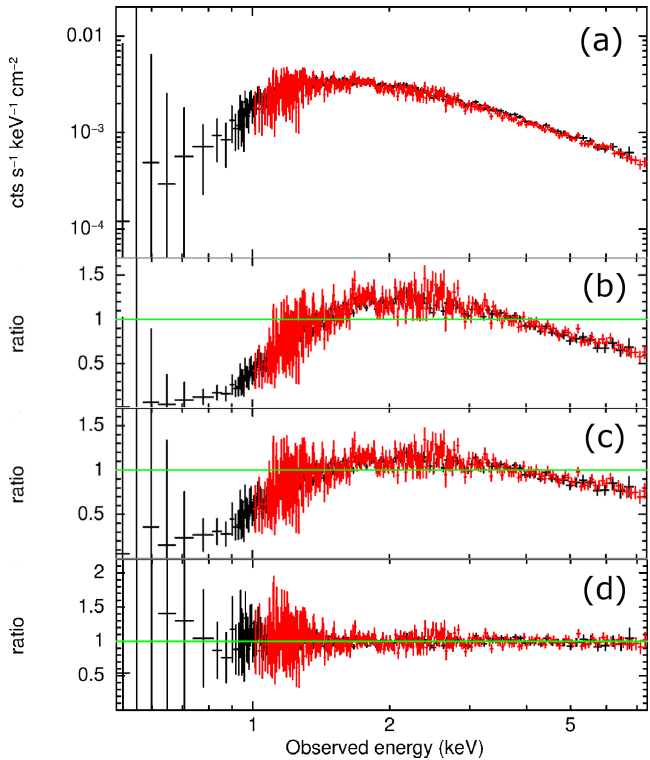}
   \caption{\emph{Chandra} MEG (black) and HEG (red) spectra
     of 3C~111. The data are binned to 4$\times$ the FWHM resolution
     and to a minimum signal-to-noise of 2 in order to emphasize the broad-band curvature of the spectrum due to the absorber column density. \emph{Panel a:} MEG and HEG spectra. \emph{Panel b:} ratio with respect to
     a power-law continuum model. \emph{Panel c:} ratio with respect
     to a power-law continuum model including the tabulated Galactic
     absorption value of $N_H = 3\times
     10^{21}$~cm$^{-2}$. \emph{Panel d:} ratio with respect to the
     best-fit model consisting of a power-law continuum, Galactic
     absorption with free column density and an Fe K$\alpha$ emission
     line.}
    \end{figure}

The MEG and HEG spectra of 3C~111 are shown in panel a of Fig.~1. We
started the spectral modeling using a power-law continuum with photon
index $\Gamma \simeq 0.9$. As
we can see from panel b of Fig.~1 this does not provide a good fit
($C/u = 8182/2220$), with the soft X-ray residuals and very flat spectral index indicating the requirement for a neutral absorption component.

We then included a neutral absorption component modeled with
\emph{tbabs} in XSPEC. This model calculates the cross
section for X-ray absorption by the interstellar medium as the sum of
the cross sections for X-ray absorption due to the gas-phase, the
grain-phase, and the molecules (Wilms et al.~2000). 

We start using a column density of $N_H = 3\times
10^{21}$~cm$^{-2}$ derived from NASA's HEASARC \emph{nh} tool\footnote{http://heasarc.gsfc.nasa.gov/cgi-bin/Tools/w3nh/w3nh.pl}. 
This is the intermediate value between the weighted average values
estimated by the Leiden/Argentine/Bonn Survey of Galactic H~I (Kalberla
et al.~2005) of $N_H = 2.91\times 10^{21}$~cm$^{-1}$ and by the Dickey
\& Lockman H~I in the Galaxy (Dickey \& Lockman 1990) of $N_H =
3.15\times 10^{21}$~cm$^{-2}$. This fit provides a high statistical improvement with
$C/u = 4145/2220$ and less flat power-law photon index of $\Gamma \simeq 1.2$.
The ratio of the spectra with respect to the power-law including the
tabulated Galactic absorption is shown in panel c of Fig.~1. We note
an improvement with respect to panel b, but the fit is still not
satisfactory.

Previous observations of the radio galaxy 3C~111 showed an excess absorption with
respect to the tabulated Galactic value of $N_H \sim 5\times
10^{21}$~cm$^{-2}$. However, it was still not possible to constrain
the origin of such absorber as local to our Milky Way or intrinsic to
3C~111 because the redshift could not be constrained (e.g., Reynolds
et al.~1998; Lewis et al.~2005; Ballo et al.~2011; Tombesi et al.~2013). 

We then included a new \emph{ztbabs} absorption component assuming
the redshift of the source. We obtained a high fit improvement with $C/u =
2264/2219$ for a column density of $N_H = (5.1\pm0.3)\times
10^{21}$~cm$^{-2}$ and a photon index of $\Gamma \simeq 1.6$. We then
left the redshift free to vary and we obtained an additional very high fit improvement
of $\Delta C/\Delta u = 25/1$, corresponding to a statistical
confidence level of 5$\sigma$. The best-fit redshift is $z =
-0.002_{-0.038}^{+0.002}$, which is consistent with being
local to our Milky Way galaxy instead of 3C~111 at a redshift of $z =
0.0485$. Considering a single neutral \emph{ztbabs} absorber component to model
both the tabulated and excess absorption we obtain a column density of
$N_H = (7.5\pm0.2)\times 10^{21}$~cm$^{-2}$ and a better constrained
redshift value of $z=-0.002^{+0.002}_{-0.008}$. The contour plots of
these parameters are shown subsequently in Fig.~4.

The main spectral features responsible for the absorption are the
photoelectric edges from O~I K at E$=$0.543~keV, Fe~I L$_1$ at
E$=$0.845~keV, Fe~I L$_2$ at E$=$0.72~keV, and Ne~I K at
E$=$0.87~keV. In particular, being the O edge so deep, the fit is
mostly driven by counts in the latter three edges. In fact, we obtain
the same results restricting the fit to energies higher that
E$=$0.7~keV. Moreover, we note that the Fe~I L$_1$ edge is significantly weaker
than the Fe~I L$_2$ edge, and therefore it will only marginally affect
the fit (e.g., Schulz et al. 2002).
We checked that we obtain the same results using different
absorption models in XSPEC, such as \emph{phabs} and \emph{wabs}, or
the detailed X-ray absorption code \emph{tbnew\_gas} (Wilms et al.~2000),
which is however still at a beta-test
version\footnote{http://pulsar.sternwarte.uni-erlangen.de/wilms/research/tbabs/}.

Looking at the ratio between the data and this model zoomed in the
E$=$6--7~keV band in Fig.~2 we observe the presence of a faint
emission line at the expected energy of the Fe K$\alpha$ fluorescence
emission line of E$=$6.4~keV. Therefore, we consider the final
best-fit model composed of a power-law continuum, a single Galactic neutral
absorber, and an Fe K$\alpha$ Gaussian emission line. The ratio of the
data with respect to the best-fit model is shown in panel d of
Fig.~1. 

Assuming a redshift of zero, i.e. local to our Milky Way, the best-fit parameters are a column density of $N_H =
(7.7\pm0.1)\times 10^{21}$~cm$^{-2}$, a power-law photon index of
$\Gamma = 1.62\pm0.02$, and an emission line at the energy of
E$=$$6.41\pm0.03$~keV with intensity $I = (1.2\pm0.7)\times
10^{-5}$~ph~s$^{-1}$~cm$^{-2}$, width $\sigma < 50$ eV, and equivalent
width EW$=$$16\pm10$~eV. The best-fit statistics is $C/u =
2232/2216$. The extrapolated absorption corrected fluxes in the energy intervals
E$=$0.5--2~keV and E$=$2--10~keV are $3.0\times 10^{-11}$
erg~s$^{-1}$~cm$^{-2}$ and $6.1\times 10^{-11}$
erg~s$^{-1}$~cm$^{-2}$, respectively.

  \begin{figure}[t!]
  \centering
   \includegraphics[width=8.5cm,height=7cm,angle=0]{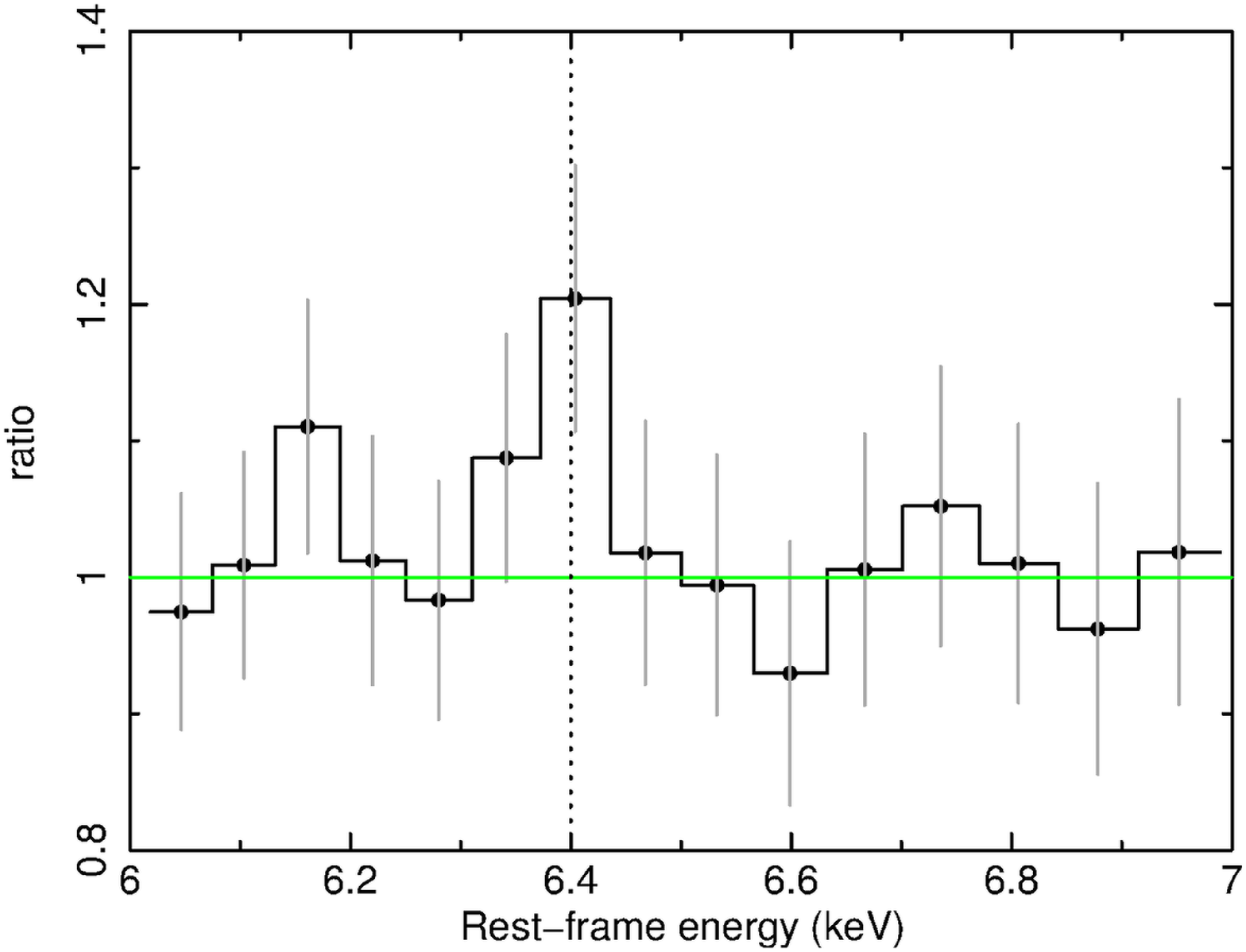}
   \caption{Ratio between the \emph{Chandra} HEG spectrum and an
     absorbed power-law model zoomed in the energy band
     E$=$6--7~keV. The vertical dotted line indicates the Fe K$\alpha$
     line at the energy of E$=$6.4~keV. The data are binned to 2$\times$ the FWHM resolution for clarity.}
    \end{figure}

The intensity of the Fe K$\alpha$ emission line is consistent with
previous estimates and the low EW is consistent with the source being
observed during a high state  (Reynolds et al.~1998; Tombesi et al.~2010b, 2011, 2012b,
2014; Chatterjee et al.~2011; Ballo
et al.~2011). We note that the full width at
half maximum (FWHM) of the Fe K$\alpha$ of FWMH$_{\text{FeK}}$$<$5,500~km~s$^{-1}$ is
consistent with the FWHM of the optical H$\alpha$ line of
FWHM$_{\text{H}\alpha}$$= 4,800\pm 200$~km~s$^{-1}$ (Eracleous \& Halpern
2003). The full width at zero intensity (FWZI) from the H$\alpha$ line
is instead much broader FWZI$= 18,400\pm3,000$~km~s$^{-1}$, but we can
not constrain this larger broadening if present also in the Fe
K$\alpha$ line due to the limited signal-to-noise in the spectrum at
these energies.

Using an \emph{XSTAR} photo-ionized absorption table with turbulent
velocity of 100~km~s$^{-1}$ and assuming the
typical ionization of the warm absorber detected in
other BLRGs of log$\xi$$\simeq$2.5~erg~s$^{-1}$~cm and a velocity consistent with
zero at the source rest-frame (e.g., Reeves et al. 2009; Torresi et
al. 2010, 2012), we estimate an upper limit of the column
density of a possible warm absorber of
$N_H$$<$$2.5\times10^{20}$~cm$^{-2}$. The fact that we do not clearly
detect a warm absorber in this source could partially be due to the
intervening absorption from our own Galaxy and/or to the fact that the
interstellar medium in this source could be hot, as observed in
3C~390.3 and 3C~120, and the low inclination estimated at $\simeq$18$^{\circ}$ from the radio jet (e.g., Torresi et
al.~2012; Tombesi et al.~2016, 2017).

\subsection{Combining Chandra and XMM-Newton spectra}

3C~111 was observed twice with \emph{XMM-Newton}, in 2001
for 45~ks and in 2009 for 120~ks. Here,
our focus is on the high-energy resolution RGS spectra in the
energy band between E$=$0.5--2.5~keV. We use both RGS1 and RGS2
detectors, and the latest pipeline data products.

  \begin{figure}[t!]
  \centering
   \includegraphics[width=8.5cm,height=7cm,angle=0]{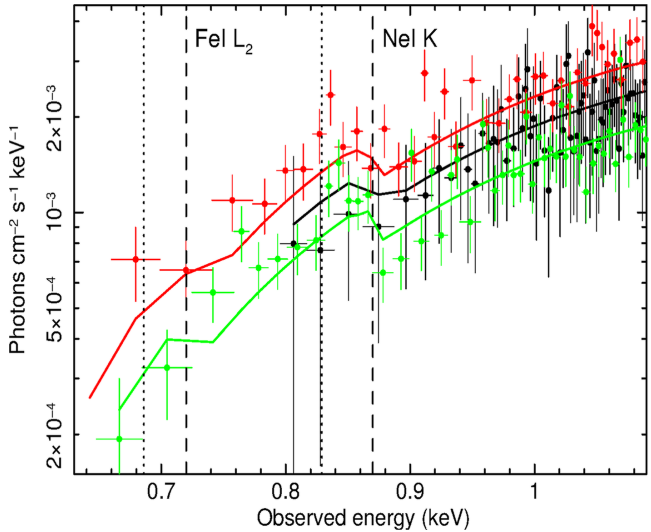}
   \caption{Grating spectra and best-fit absorbed power-law model
     zoomed in the E$=$0.6--1.1~keV band. The \emph{Chandra} MEG
     (black) is binned to 2$\times$ the FWHM resolution and to a
     minimum signal-to-noise of 2 for clarity. The \emph{XMM-Newton}
   RGS2 spectra taken in 2001 (red) and 2009 (green) are binned to a
   minimum signal-to-noise of 5 for clarity. The MEG data below
     E$=$0.8~keV are not shown here because the large error bars would
     make the comparison with the RGS less clear. The vertical dashed and
   dotted lines indicate the edge energies for Fe~I L$_2$ at E$=$0.72~keV and Ne~I K
   at E$=$0.87~keV for a redshift of zero and intrinsic to 3C~111, respectively.}
    \end{figure}

First, we analyzed the RGS data alone using a model composed of a
Galactic absorbed power-law continuum with extra cold absorption. 
Given the limited energy band of the RGS, we consider a power-law
continuum of $\Gamma = 1.6$ as estimated from previous broadband
spectral analyses of these observations (Lewis et al.~2005; Ballo et
al.~2011). Also in this case we find an extra absorption column
density of $N_H \simeq 5\times 10^{21}$~cm$^{-2}$, with a redshift
consistent with zero with an uncertainty of $\Delta z \simeq
0.02$. Therefore, also the RGS data alone favor an absorber in the Milky Way, with a
confidence level of $\sim 4 \sigma$. 

We then performed a combined fit of the \emph{Chandra} HETG and
\emph{XMM-Newton} RGS spectra allowing for free power-law normalizations and
cross-normalizations between observations and instruments,
respectively. When fitting with a redshift fixed to the one of 3C~111
($z = 0.0485$) or free to vary for the excess absorber, we find that a
value consistent with zero is favored at $> 6 \sigma$ ($\Delta C / \Delta
u = 43/1$). 

Considering only one absorber, we estimate a power-law photon index of $\Gamma
= 1.60\pm0.01$, a total Galactic absorption column density of $N_H =
(7.4\pm0.1)\times 10^{21}$~cm$^{-2}$, and a redshift of $z =
-0.002^{+0.003}_{-0.002}$. We consider this total time-averaged column density estimate to
be the best-fit value from our analysis. 
The grating spectra and the best-fit absorbed power-law model zoomed in
the E$=$0.6--1.1~keV band for the \emph{Chandra} MEG and the
\emph{XMM-Newton} RGS2 taken in 2001 and 2009 are shown in Fig.~3. Instead, the contour plots of the column density
with respect to the absorber redshift for the fits using the \emph{Chandra} HETG and
\emph{XMM-Newton} RGS spectra alone, and both combined, are shown in Fig.~4. 

Given that the three observations were performed over a time-scale of about a decade, we also performed a multi epoch spectral fit leaving the column density and redshift free to vary. We obtain that the redshift is always consistent with zero, but the column density shows a possible increase from $N_H = (7.0\pm0.02)\times 10^{21}$~cm$^{-2}$ for the \emph{XMM-Newton} observation in 2001 to a value of $N_H = (7.5\pm0.02)\times 10^{21}$~cm$^{-2}$ for the \emph{XMM-Newton} and \emph{Chandra} observations of 2009 and 2014, respectively. Although very marginal, this increase in column density is in line with the reported temporal variability in the strength and
profile of the 4.83~GHz H$_2$CO absorption line in radio observations, which show that the cloud complex may 
contain inhomogeneities even on sub-parsec scales (e.g., Marscher, Moore \&
Bania 1993; Moore \& Marscher 1995).

We checked also for variable O, Ne, and Fe
abundances using the \emph{tbvarabs} model in XSPEC (Wilms et al.~2000). The abundances
are consistent with the Solar values within the uncertainties and we are able to
place only lower limits of $A_{\text{O}} \ge 0.5$, $A_{\text{Ne}} \ge
0.9$, and $A_{\text{Fe}} \ge 0.9$, respectively. 

Finally, we performed a consistency check analyzing also the
\emph{XMM-Newton} EPIC-pn spectra simultaneous with the RGS
observations performed in 2001 and 2009. The EPIC-pn has the
highest sensitivity in the energy interval E$=$0.5--10~keV. 
However, we note that the resolving power of the EPIC-pn of $E/\Delta
E$$\simeq$5--10 at the energies of E$=$0.5--1.0~keV is very limited compared to
$E/\Delta E \simeq 300$ for the RGS and $E/\Delta E \simeq 600$ for
the HETG, respectively. Moreover, there are significant
cross-calibration uncertainties of both the shape and normalization of the effective area between the EPIC-pn and RGS.

We extracted the source and background spectra from 40 arcsec circular
regions on the detectors and applyed standard reduction techniques. We
consider the absorbed power-law continuum model in the whole energy interval of E$=$0.5--10~keV. 
The column densities are estimated to be $N_H =
(0.61\pm0.04)\times 10^{21}$~cm$^{-2}$ and $N_H =
(0.73\pm0.03)\times 10^{21}$~cm$^{-2}$, respectively.
We can place an upper limit on the redshift of the absorber of $z <
0.025$, which is lower than the one of 3C~111 of $z = 0.0485$. These
values are in agreement with the much more accurate results derived from
the analysis of the RGS and HETG spectra.

  \begin{figure}[t!]
  \centering
   \includegraphics[width=8.5cm,height=7cm,angle=0]{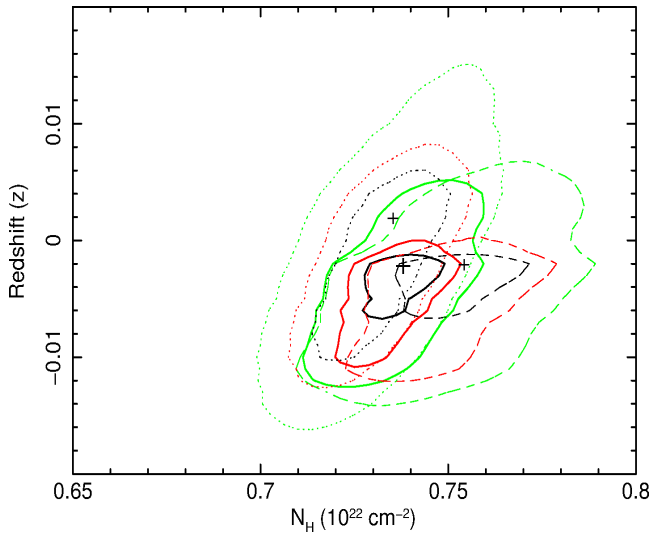}
   \caption{Contour plots comparing the column density with respect to
     the cosmological redshift of a single neutral
       absorber component modeling both the tabulated and the
       excess Galactic absorption. The contours
     refer to \emph{Chandra} HETG (dashed), \emph{XMM-Newton} RGS
     (dotted), and their combined fit (solid). The crosses indicate
     the best-fit values. The contours refer to confidence levels of
     68\% (black), 90\% (red), and 99\% (green), respectively.}
    \end{figure}

\section[]{Discussion}

The radio galaxy 3C~111 was known from previous observations to show
excess neutral absorption with respect to the column density of $N_H = 3\times
10^{21}$~cm$^{-2}$ estimated from 21~cm radio surveys of H~I in the
Galaxy (Dickey \& Lockman 1990; Kalberla et al.~2005). 
However, it was still not possible to constrain the origin of such
absorber as local to our Milky Way or intrinsic to 3C~111 ($z = 0.0485$) because the
redshift of this component could not be constrained (e.g., Reynolds
et al.~1998; Lewis et al.~2005; Tombesi et al.~2013; Ballo et
al.~2011). The analysis of the \emph{Chandra}
HETG and \emph{XMM-Newton} RGS spectra of 3C~111 allowed us
to constrain the excess absorption as due to intervening gas in the
Milky Way and to estimate a total time-averaged column density of $N_H =
(7.4\pm0.1)\times 10^{21}$~cm$^{-2}$. What is the origin of the extra
Galactic neutral absorption of $N_H = 4.4\times 10^{21}$~cm$^{-2}$
with respect to the atomic H~I value?

3C~111 is located at a relatively low latitude (b$=$$-8.8^{\circ}$)
with respect to the Galactic plane and it is known to lie behind the
giant Taurus molecular cloud, which is the nearest large star-forming region in our
Galaxy, at an estimated distance of about 200~pc from us (Ungerer et
al.~1985; Ungerechts \& Thaddeus 1987). Reynolds et al.~(1998) suggested that this cloud
complex may cause variations between the actual Galactic absorption
along our line of sight and that inferred from 21~cm radio measurements due to inhomogeneities or the
presence of molecular hydrogen.

Temporal variability in the strength and
profile of the 4.83~GHz H$_2$CO absorption line in radio observations show that the cloud complex
contains inhomogeneities on sub-parsec scales, and that between 30 and 100
clumps may lie along the line of sight to 3C~111 (Marscher, Moore \&
Bania 1993; Moore \& Marscher 1995). A column density change
is also marginally evident comparing the X-ray spectra performed a
decade apart in sec.~2.1. These inhomogeneities could in
principle partially explain the discrepancy in column density
estimates, but 3C~111 does not appear to be blocked by one or more
particularly dense clumps (Reynolds et al.~1998). 

Molecular hydrogen along our line of sight to 3C~111 associated with the Taurus molecular cloud may
explain the excess. In fact, this gas would not contribute to the
21~cm emission, but the metals/dust associated with it will have
opacity in the X-ray band. If about 60\% of the gas along the line of
sight to 3C~111 is in molecular form rather than atomic, the
discrepancy between the 21~cm radio measurements and the X-ray could be resolved.

Pineda et al~(2010) show a linear relationship between CO and
extinction $A_V$ in the Taurus molecular cloud. Their results allow to
estimate the column of molecular hydrogen H$_2$ with respect to the
extinction: $N_{H_2} = 9.4\times 10^{20} A_V$. The extinction along
the line of sight to 3C~111 is\footnote{From
  https://ned.ipac.caltech.edu/} 4.53, which gives a molecular column
density of $N_{H_2} \simeq 4.3\times 10^{21}$. This value is indeed
consistent with the excess absorption with respect to the atomic H~I. This is also consistent
with the more general relation between the total column density of H~I
plus $H_2$ reported in Bolatto et al.~(2013) of $N_H \simeq 1.9\times 10^{21} A_V$~cm$^{-2}$,
which provides an estimate of $N_H \sim 8\times 10^{21}$~cm$^{-2}$
toward 3C~111. Finally, this is overall in agreement with the molecular hydrogen column density
of $N_{H_2} \simeq 4.5 \times 10^{21}$ molecules~cm$^{-2}$
estimated from the interstellar CO emission line measurement derived by Bania, Marscher \&
Barvainis (1991) toward 3C~111. However, we note that the cloud in
front of 3C~111 is translucent rather than opaque, and the H$_2$ to CO
ratio is better constrained for opaque clouds.

\section[]{Conclusions}

In this paper we show that the cold absorption detected in the X-ray
band toward 3C~111 is indeed of Galactic origin and it is very likely
due to a combination of atomic and molecular gas.
Therefore, we recommend to use the total Galactic column density here
estimated to be $N_H = (7.4\pm0.1)\times 10^{21}$~cm$^{-2}$ when studying
the radio galaxy 3C~111. For instance, a recent spectral energy
distribution study of the radio, optical, IR, and X-ray knots in the
extended jet of 3C~111 was found to be dependent on the assumed
Galactic column (Clautiche et al.~2016). 
On the other hand, the presence of these complexities toward 3C~111
make this object a very promising background source for
multiwavelength studies of the characteristics of the atomic and
molecular gas in the Taurus molecular cloud, which is the closest large star-forming region in our
Galaxy (e.g., Ungerer et al.~1985; Marscher, Moore \& Bania 1993;
Moore \& Marscher 1995; G{\"u}del et al.~2007). 
The synergy between future deeper high-energy resolution X-ray
observations and multi-wavelength campaigns will allow us to
investigate in more details the characteristics of the complex multi-phase
medium in the Milky Way, and constrain the composition, elemental abundance, distribution, and the presence of dust.

\acknowledgments

F.T. thanks A.~D. Bolatto and A.~P. Marscher for the useful comments.
F.T. and C.S.R. acknowledges support for this work by the National Aeronautics and Space
Administration (NASA) through Chandra Award Number GO4-15103A issued
by the Chandra X-ray Observatory Center, which is operated by the
Smithsonian Astrophysical Observatory for and on behalf of NASA under
contract NAS8-03060. F.T. acknowledges partial support by the
Programma per Giovani Ricercatori - anno 2014 ``Rita Levi
Montalcini''. E.B. is supported by the European Union’s Horizon 2020 research and innovation programme under the Marie Sklodowska­Curie grant agreement no. 655324, and by the I­CORE program of the Planning and Budgeting Committee (grant number 1937/12).

\end{document}